\documentclass[reprint,3p,times,onecolumn,12pt]{elsarticle}
\pdfoutput=1 
\usepackage{geometry}
\usepackage{mathtools}
\usepackage{graphicx}
\usepackage{latexsym}
\usepackage{amssymb}
\usepackage{amscd,amssymb}
\usepackage[arrow,matrix]{xy}
\usepackage{graphicx}
\usepackage{epstopdf}

\biboptions{sort&compress}

\usepackage{ulem}
\usepackage{setspace}

\usepackage{amscd,amsmath,amsthm,amssymb}
\usepackage{amsmath,amssymb}
\usepackage[colorlinks=true,bookmarks=false,citecolor=blue,urlcolor=blue]{hyperref} %pdflatex

\newcommand{\bea}{\begin{eqnarray}}
\newcommand{\eea}{\end{eqnarray}}
\newcommand{\bes}{\begin{subequations}}
\newcommand{\ees}{\end{subequations}}

\begin{document}\title{Engineering Optical Rogue Waves and Breathers in a Coupled Nonlinear Schr\"odinger System with Four-Wave Mixing Effect}
	
	\author[ks1,ks2]{\large K. Sakkaravarthi \corref{cor}}
	\author[rb]{\large R. Babu Mareeswaran}
	\author[tk]{\large T. Kanna\corref{cor}}
	
	\address[ks1]{Department of Physics, National Institute of Technology, Tiruchirappalli -- 620 015, Tamil Nadu, India}
	\address[ks2]{Centre for Nonlinear Dynamics, School of Physics, Bharathidasan University, Tiruchirappalli -- 620 024, India}
	\address[rb]{\it PG and Research Department of Physics, PSG College of Arts and Science, Coimbatore--641 014, India}
	\address[tk]{\it Nonlinear Waves Research Laboratory, PG and Research Department of Physics,\\ Bishop Heber College (Affiliated to Bharathidasan University), Tiruchirapalli--620 017, Tamil Nadu, India\vspace{-1.0cm}}
	
	\cortext[cor]{Corresponding authors. \newline Email address: ksakkaravarthi@gmail.com; babu\_nld@rediffmail.com; kanna\_phy@bhc.edu.in}
	
	\journal{\bf Physica Scripta}
	\date{}
 
\begin{abstract}
	We consider a coherently coupled nonlinear Schr\"odinger equation with modulated self-phase modulation, cross-phase modulation, and four-wave mixing nonlinearities and varying refractive index in anisotropic graded index nonlinear medium. By identifying an appropriate similarity transformation, we obtain a general localized wave solution and investigate their dynamics with a proper set of modulated nonlinearities. In particular, our study reveals different manifestations of localized waves such as stable solitons, Akhmediev breathers, Ma breathers, and rogue waves of bright, bright-dark, and dark-dark type and explores their manipulation mechanism with suitably engineered nonlinearity parameters. We have provided a categorical analysis with adequate graphical demonstrations.
	
	\begin{keyword}{Coherently coupled nonlinear Schr\"odinger equation; Similarity transformation; Solitons; Akhmediev breather; Ma breather; Rogue waves.\newline}
		------------------------------------------------------------------------------------------------------------------------\newline
\noindent {\large Journal Reference: {\it Physica Scripta} {\bf 95} (2020) 095202.\newline
	 \url{https://doi.org/10.1088/1402-4896/aba664}}
	\end{keyword}
\end{abstract}

\maketitle

\setstretch{1.25}

\section{Introduction}
Dynamics of nonlinear systems features several interesting phenomena that find multifaceted applications in different fields of science, engineering, and technology through a systematic understanding of the respective mathematical models. The existence of various nonlinear coherent structures associated with such models is of considerable physical significance \cite{wiley}. Among several types of localized nonlinear waves, solitons receive profound interest due to their remarkable stability nature and intriguing collision dynamics. For the past few decades, studies on optical/atomic solitons have shown promising results in theoretical aspects and in their corresponding experimental realization through which a number of phenomena have been demonstrated in various fields \cite{kiv-book}. Apart from the solitons, the amusing nonlinear coherent structure, namely rogue waves that appear from nowhere and disappear without a trace \cite{akm}, have attracted significant attention during the past ten years or so from theoretical and experimental perspectives \cite{Nat-OptRogue,rogue-exp1,rogue-exp2,rogue-exp3}. These are because of their more frequent appearance in deep sea and due to their experimental observation in information transfer through optical fibers, atomic condensates, and different dynamical systems in recent years. These rogue waves (or freak/monster/killer/extreme/abnormal waves) were first observed in deep ocean water extreme amplitude waves with a significant height that is a few times higher than the average wave crests \cite{rogue-book}. It is difficult to conduct experiments in the deep ocean to understand them due to the dangerous conditions. However, observations from the replicated laboratory experiments and cross-disciplinary investigations (for example, in optical and atomic systems and so on) can be adopted for realizing their complete dynamical evolution as well as prediction mechanism. Such rogue waves are a special case of breathers that are another type of coherent structure oscillating periodically in space or time and display periodic variation in their amplitude during propagation \cite{rogue-book}. These breathers have also received significant importance in recent years. Theoretically, these nonlinear coherent structures can be well described by the ubiquitous nonlinear Schr\"odinger (NLS) equation \cite{akm}, its variants and similar models. Though they are initially observed as extreme waves in the oceans, now a days they appear in a wider range of fields such as fiber optics as extreme amplitude pulses, optical cavities as high amplitudes, laser outputs as extreme pulses, Bose-Einstein condensates (BEC) as high atomic concentration, etc. \cite{rogue-book,Nat-Rev19}.

Beyond these localized waves, several types of nonlinear waves arise in various nonlinear dynamical models under different circumstances. We do not pursue/discuss them here considering the objective and length. Indeed multicomponent nonlinear waves show promising characteristics than their single-component/scalar counterparts and hence coupled nonlinear systems are under continuous exploration. 
It is clear from the literature that there exist plenty of multicomponent mathematical models, especially differential/difference equations, describing the dynamics of various physical, chemical, biological, engineering, and even medical systems and finance. Appropriate analysis of these models provides an understanding of the underlying system. Those analyses can be done by constructing explicit solutions using/developing analytical (semi-analytical) methods, which may be possible for certain classes of equations, namely integrable models. However, most of the natural systems are non-integrable, forming another category, namely, exactly non-solvable models that require strong computational tools in addition to some symmetry and approximate methods \cite{Boris-book}.

Keeping the above perspectives in mind, in this work, we report the characteristic dynamics of certain localized waves in an inhomogeneous nonlinear optical model \cite{kiv-book,Boris-book}. To be precise, we take the following coherently coupled nonlinear Schr\"odinger (CCNLS) equation in dimensionless form:
\bes\bea
i\frac{\partial q_1}{\partial z} - \frac{\partial^2 q_1}{\partial t^2} - \sigma(z)\left( |q_{1}|^2+2 |q_{2}|^2\right)q_{1}-\sigma(z) q_2^2q_{1}^*+\Omega(z,t)q_1=0,\\
i\frac{\partial q_2}{\partial z} - \frac{\partial^2 q_2}{\partial t^2} - \sigma(z)\left(2 |q_{1}|^2+|q_{2}|^2\right)q_{2}-\sigma(z) q_1^2q_{2}^*+\Omega(z,t)q_2=0,
\eea\label{gen-2ccnls}\ees
where the derivatives with respect to $z$ and $t$ represent the propagation direction and transverse coordinate, respectively. 
The above equation is referred as coherently coupled nonlinear Schr\"odinger equation with varying nonlinearity $\sigma(z)$ governs the co-propagation of two complex envelope optical modes $q_j(z,t),~j=1,2,$ in a medium like optical fiber with modulated parabolic refractive index profile $\Omega=\frac{1}{2}\mu(z)t^2$ for the linear part of the refractive index. To be precise, one can express the effective refractive index of the optical system as $n(z,t)=n_0+n_1 \mu(z) +n_2 \sigma(z) I(z,t)$, where $n_0$ denotes constant refractive index of the medium while $n_1$ and $n_2$ represent the modulated linear and nonlinear (Kerr type) contribution of refractive index, and $I$ is the total intensity of the optical wave \cite{kiv-book}. In one way, the above type of model can appear in cubical nonlinear (Kerr) medium with graded refractive index \cite{rogue-exp2,rogue-exp3}. Also, equation (\ref{gen-2ccnls}) contains coherent coupling nonlinearity resulting from four-wave mixing effect in addition to the standard incoherent nonlinearities self-phase modulation (SPM) and cross-phase modulation (XPM). The above equation is non-integrable while its homogeneous version turns out to be integrable for which and for several similar models various nonlinear wave solutions including solitons, periodic waves, breathers, rogue waves, etc. have been obtained by using various methods \cite{Park,tkjpa,ksjpa,ksjmp,tkpla16,NLD-15,NLD-15a,NLD-20,AML-20a,pre2015,pre2015a,AML-20,RSPA-17,EPL-19} and one version is also referred as pair-transition-coupled NLS model \cite{pre2015,pre2015a,RSPA-17,AML-20}. Further, there exist several studies on certain scalar inhomogeneous models and their related coupled systems exploring the dynamics of different nonliner waves in both one- and higher-dimensions, to mention a few  \cite{refsug1,refsug2,refsug3,refsug4,refsug5,refsug6}. However, the present inhomogeneous system requires an authoritative investigation, which motivated us to proceed with the analysis for a category of localized structures in this report. Though localized waves arise in different contexts from hydrodynamics to atomic condensates, control strategies of such waves to understand their transitions are still difficult/less (particularly, engineering the extreme waves) due to limited experimental implementation and a few tools are being developed in recent years \cite{NatCom-rogue-2019,PRX-rogue-2019,PRE-rogue-2018,SciRep-Akm-12,RSPA,Rogue-review-2020}. Thus, we devote ourselves to manipulating such localized nonlinear waves through explicit solutions with inhomogeneity parameters and will concentrate on exploring their dynamical features.

We provide the essential mathematical tool of similarity transformation briefly along with the estimated nonlinearities in Sec. \ref{sec-simi}. We construct the inhomogeneous localized nonlinear wave solution in Sec. \ref{sec-solution} and discuss the impact of chosen modulated nonlinearities in each type of nonlinear waves and the importance of available control parameters in Sec. \ref{inhomo-solution}, where we have also given certain possible future directions along this study. Finally, we summarize the important outcomes in Sec. \ref{sec-conclusion}. 

\section{Similarity Transformation \& Nonlinearity Management}\label{sec-simi}
Among various methodologies to solve nonlinear models, similarity transformation occupies a prominent place due to its simple adaptability/applicability in executing the objectives. To begin with, we consider the following form of solution to Eq. (\ref{gen-2ccnls}):   
\bea
q_j(z,t) =\epsilon_1 \sqrt{\sigma(z)}~Q_j(Z(z),T(z,t))~\text{exp}[i\xi(z,t)],\quad j=1,2,
\label{simi}
\eea
where $\epsilon_1 \sqrt{\sigma(z)}$ determines the amplitude, while $\xi(z,t)$ is the phase and $T(z,t)$ and $Z(z)$ are the similarity variables, the explicit form of all these variables has to be determined in a systematic way. On the substitution of Eq. (\ref{simi}), the model (\ref{gen-2ccnls}) reduces to a set of coupled equations of $Q_j(Z,T)$ in the form
\bes\bea
&iQ_{1Z} - Q_{1TT}- (|Q_1|^2+2|Q_2|^2)Q_1 -  Q_2^2Q_1^*=0,\\
&iQ_{2Z} - Q_{2TT}-(2|Q_1|^2+|Q_2|^2)Q_2- Q_1^2Q_2^*=0,
\eea\label{ccnls}\ees
along with a set of constraint relations in terms of varying nonlinearity and arbitrary constants $\epsilon_1$ and  $\epsilon_2$ as given below. 
\bes\bea
&&\xi(z,t) = \frac{1}{4}\frac{\sigma_z}{\sigma}t^2 +  \epsilon_1^2 \epsilon_2 {\sigma} t + \epsilon_2^2 \epsilon_1^4\int {\sigma}^2 dz,\\
&&T(z,t) = ~\epsilon_1 \left({\sigma} t + 2 \epsilon_2 \epsilon_1^2\int {\sigma}^2 dz\right),\\
&&Z(z) = \epsilon_1^2 \int {\sigma}^2 dz, \\
%&&\rho(z)=\epsilon_1 \sqrt{\gamma(z)},\\
&&\mu(z)=\frac{\sigma_{zz}}{2\sigma}-\frac{\sigma_z^2}{\sigma^2},\label{potF}
\eea\label{cons}\ees
Equation (\ref{ccnls}) is nothing but an integrable CCNLS type model and its solutions will indirectly give the corresponding solutions of (\ref{gen-2ccnls}) through Eqs. (\ref{simi}) and (\ref{cons}). As mentioned in the introduction, studies on different nonlinear waves (especially solitons, breathers, rogue waves) of Eq. (\ref{ccnls}) and its variants are abundant including an interesting energy-switching collisions of single-double-hump solitons \cite{tkjpa} and higher order rogue waves \cite{AML-20,NLD-20,RSPA-17,NLD-15,NLD-15a,EPL-19}.

\begin{figure}[!pht]	\centering\includegraphics[width=.32\linewidth]{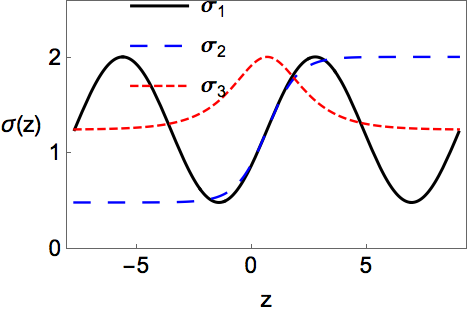}~\includegraphics[width=.33\linewidth]{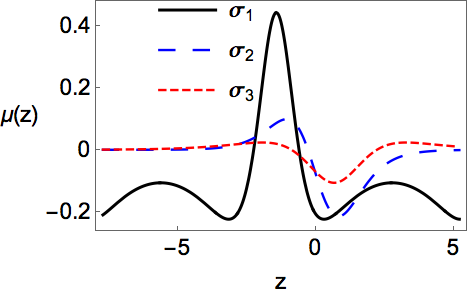}~\includegraphics[width=.35\linewidth]{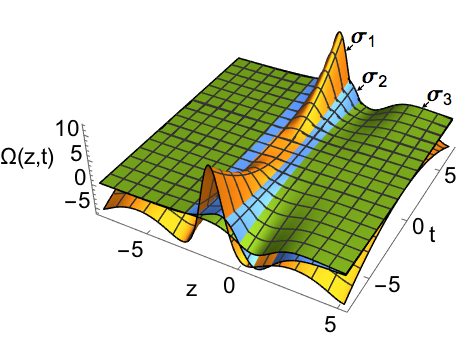}\\{\hfill (a) \hfill \hfill (b) \hfill \hfill (c) \hfill} 
	\caption{Nature of three different nonlinearities $\sigma(z)$ (a) and respective modulation in refractive index profile $\mu(z)$ (b) and $\Omega(z,t)$ (c) for $b_0=1.25$, $b_1=0.76$, $b_2=0.75$, and $b_3=-0.5$.}
	\label{fig-non-pot}
\end{figure}

In the present study, we focus on an interesting solution structure supporting multi-wave evolution and will demonstrate the impact of varying nonlinearity of different forms that drive the nature of the refractive index of a harmonic oscillator. The nonlinearity modulations considered in this work are (i) periodic, (ii) step-like, and (iii) bell-type nonlinearities, respectively in the form 
\bes \bea 
&&\sigma_1(z)=b_0+b_1 ~\text{sin}(b_2 z+b_3),\\ 
&&\sigma_2(z)=b_0+b_1 ~\text{tanh}(b_2 z+b_3), \\ 
&&\sigma_3(z)=b_0+b_1~ \mbox{sech}(b_2 z+b_3), 
\eea \label{nonlinearity} \ees 
where $b_0$, $b_1$, $b_2$, and $b_3$ are arbitrary real constants. Further, the mutual dependency of the refractive index profile on this nonlinearity can be given in an explicit form through Eq. (\ref{cons}d). For illustrative purpose, we have shown their modulation with respect to $z$ in Fig. \ref{fig-non-pot} and their impact on the designated nonlinear waves will be analyzed categorically. 

\section{Localized Wave Solution}\label{sec-solution}
In order to analyze the influence and respective applications of nonlinearity variation in the system (\ref{gen-2ccnls}), we need to construct its solution. For this purpose, we adopt the solution methodology proposed by Zhao {\it et al} for another version of the CCNLS model in Ref. \cite{pre2015} where Darboux transformation method was employed. It is well-known that the Darboux transformation is one of the efficient analytical tools to obtain a variety of nonlinear wave solutions of equations which admit Lax pair. We have computed the exact solution for the present CCNLS equation (\ref{gen-2ccnls}) through an appropriate scaling and the derived similarity transformation (\ref{simi}). An important reason for considering this type of solution is by virtue of only two arbitrary parameters rich dynamical features can be unearthed appropriately. 

Without going into much details, by following Ref. \cite{pre2015} and the similarity transformation given above in Sec. \ref{sec-simi}, we obtain a general localized wave solution to equation (\ref{gen-2ccnls}) which can be written in an explicit form as 
\bes \bea
%Q_1= Q_2+s\exp{[2i s^2Z]}, \qquad Q_2= \frac{2a}{|P_1|^2+|P_2|^2} P_1 P_2^*,
&&q_1=  \epsilon_1 \sqrt{\sigma}\left(\frac{2a P_1 P_2^*}{|P_1|^2+|P_2|^2} +s~e^{2i s^2Z}\right) e^{i\xi}, \\ 
&&q_2= \epsilon_1 \sqrt{\sigma}\frac{2a P_1 P_2^*}{|P_1|^2+|P_2|^2} ~e^{i\xi},
\eea 
where 
\bea 
&&P_1=  \frac{1}{\tau}\left({\sqrt{a+\tau}}~\Psi_1-{\sqrt{a-\tau}}~\Psi_2\right),\quad P_2= \frac{1}{\tau}\left({\sqrt{a+\tau}}~\Psi_2-{\sqrt{a-\tau }}~\Psi_1\right) e^{2i s^2Z},\\
&&\Psi_1=  \exp{\left[\tau T-i\left( s^2+2 a  \tau \right) Z\right]},~  
\Psi_2= \exp{\left[-\tau T-i\left( s^2-2 a \tau\right) Z\right]},~
\tau=\sqrt{a^2-s^2}\neq 0. 
\eea \label{solution} \ees  
Further, the form of $Z(z)$, $T(z,t)$ and $\xi(z,t)$ appearing in the above solution is as given in Eq. (\ref{cons}). Thus our solution has four arbitrary real parameters $\epsilon_1$, $\epsilon_2$, $s$ and $a$ in addition to the freedom to adopt any nonlinearity function. Especially, the latter two arbitrary real parameters $s$ and $a$ will play a crucial role in manipulating the above solution into different varieties of nonlinear waves such as solitons, breathers, and rogue waves. Compared to many localized wave solutions, this is more interesting because of its multifaceted nature merely with two arbitrary parameters. For example, a single arbitrary real parameter in a scalar NLS solution results in the formation of breathers and rogue waves \cite{SciRep-Akm-12,KhawajaPLA14}, but they are of bright type only. However, the present solution (\ref{solution}) is much more general and provides bright, gray, dark, bright-dark, gray-dark, and dark-dark type breathers and rogue waves in addition to the standard bright solitons. In the forthcoming sections, we will explain and discuss all such solutions as well as their modulation effects.

\section{Dynamics of Localized Waves under Inhomogeneous Nonlinearities}\label{inhomo-solution}
As pointed out above, the obtained inhomogeneous wave solution (\ref{solution}) comprises various types of localized structures ranging from the standard solitons to Akhmediev as well as Kuznetsov-Ma type breathers and rogue waves. One can rewrite the general solutions (\ref{solution}) in terms of hyperbolic and periodic functions and can derive explicit expression to each nonlinear wave structures which are given below for different possible combinations of these two arbitrary parameters `$s$' and `$a$'. Through this investigation, based on the nature of $\tau^2$ and $s/a$ defined by the appropriate choices of `$s$' and `$a$' parameters, we have categorized the resulting wave patterns and summarized them in Table \ref{table1}. We investigate the consequences of modulated nonlinearities in each of them in this section one by one, except for the singular structures case (viii). In cases (vi) and (vii), the values of $\tau^2$ are non-zero, but very small and close to zero ($\tau^2\neq 0$ and $-1<<\tau^2<<1$). Dark localized structure in the $q_1$ component comprises either dark or gray type profile for which the lowest intensity of the dip is zero for the former while non-zero in the latter. Importantly, these nonlinear localized structures can be obtained even for a fixed ``$a$" with suitably changing ``$s$" value, such that this ``$s$" shall be referred as ``switching parameter" as it switches among solitons, Akhmediev and Ma breathers, and rogue waves. 
\begin{table}[h]
	\caption{Formation of different nonlinear wave structures based on the $s$ and $a$ parameters.}
	\begin{center}	\begin{tabular}{c c c c c c c l} \label{table1}
			Case & $a$ & $s$ & $\tau^2$ & $s/a$ & $q_1$ & $q_2$ & Wave nature\\ \hline 
			(i) & $\pm$ & $0$ & $+$ & 0 & Bright & Bright & Solitons\\ 
			(ii) & $\pm$ & $+$ & $-$ & $\pm$ & Bright & Dark & Akhmediev Breathers\\ 
			(iii) & $\pm$ & $-$ & $-$ & $\mp$ & Dark & Dark & Akhmediev Breathers\\ 
			(iv) & $\pm$ & $\pm$ & $+$ & $+$ & Bright & Dark & Ma Breathers\\ 
			(v) & $\pm$ & $\mp$ & $+$ & $-$ & Dark & Dark & Ma Breathers\\ 
			(vi) & $\pm$ & $\pm$ & $\pm$ & $+$ & Bright & Dark & Rogue waves\\ 
			(vii) & $\pm$ & $\mp$ & $\pm$ & $-$ & Dark & Dark & Rogue waves\\ 
			(viii) & $\pm $ & $\pm a$ & $0$ & $*$ & $*$ & $*$ & Singular structures\\ \hline
	\end{tabular} \end{center}
\end{table}

\subsection{Inhomogeneous Solitons}
\begin{figure}[h]
	\centering\includegraphics[width=0.25\linewidth]{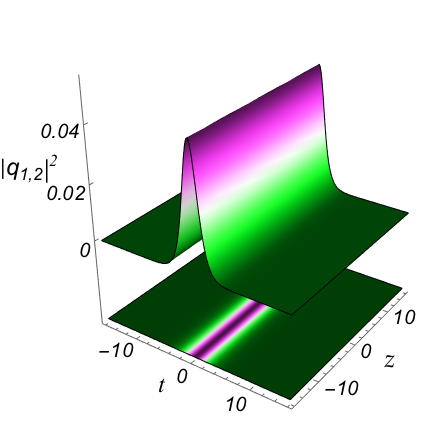}\includegraphics[width=0.25\linewidth]{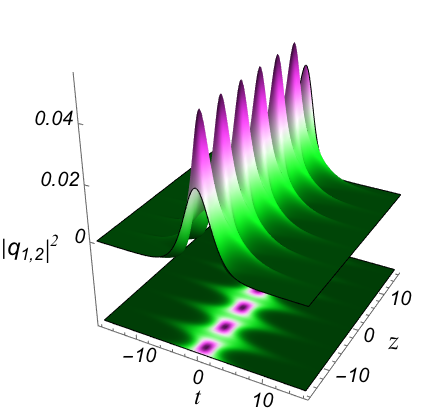}\includegraphics[width=0.25\linewidth]{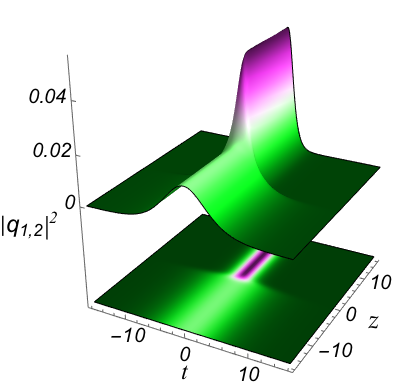}\includegraphics[width=0.25\linewidth]{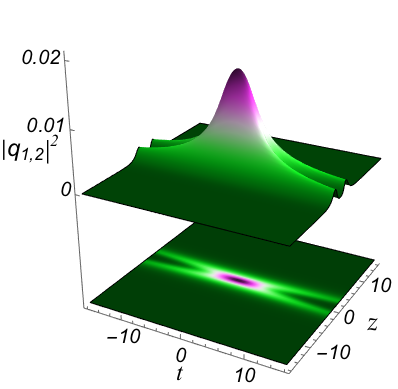}\\ {\hfill\qquad (a) \hfill\qquad (b) \hfill \qquad (c) \hfill \qquad\qquad(d) \hfill} 
	\caption{Dynamics of stationary solitons in inhomogeneous optical fiber for $s=0$ and $a=1.0$ with (a) constant, (b) periodic $\sigma_1$, (c) kink-like $\sigma_2$, and (d) bell-type $\sigma_3$ nonlinearities given in (\ref{nonlinearity}), showing the (a) stable soliton, (b) periodic breather, (c) amplification with compression, and (d) rogue-wave type exciton, respectively. Here the nonlinearity parameters are chosen as (a) $b_0 = 2.0$, $b_1 =b_2 = 0.0$, (b) $b_0 = 1.5$, $b_1 = 0.75$, $b_2 = 1.2$, (c) $b_0 = 1.5$, $b_1 = 0.75$, $b_2 = 1.2$, and (d) $b_0 = 0.0$, $b_1 = 0.75$, $b_2 = 1.2$ with other values fixed as $\epsilon_1 = 0.25$, $\epsilon_2 = 0.05$, and $b_3=0.02$.} 
	\label{fig-soliton}
\end{figure}

A clear analysis of solution (\ref{solution}) reveals that both components $q_1$ and $q_2$ will have the same type of stable localized waves, which are nothing but stationary bright solitons. This is possible when the switching parameter is absent ($s=0$) which leads to $\tau>0$, and it can be written in a convenient hyperbolic form as
\bea
&&q_1=q_2= a\epsilon_1 \sqrt{\sigma} ~ \mbox{sech}\left[2a \epsilon_1 \left({\sigma} t + 2 \epsilon_2 \epsilon_1^2\int {\sigma}^2 dz\right)\right] ~e^{i(\xi-4a^2 Z)}.  \label{sol-solution} \eea 
As a special case of this solution, first, we briefly discuss the homogeneous case with constant nonlinearity $\sigma(z)=\mbox{constant}$. 
If we check out such a homogeneous soliton solution, it is evident that there is not much freedom in its dynamics except the amplitude and width controlled by $a$. On increasing the `$a$' parameter, soliton amplitude increases (directly proportional) while its width decreases as it is inversely proportional to $a$, but do not propagate and remains stationary over time. This is one limitation of the obtained solution as it neither shows any effect of nonlinear coherent coupling nor the multi-component nature of the system, which usually portrays a rich dynamics and collisions. For example, the general soliton solution directly constructed with another method namely Hirota bilinearization method revealed a variety of soliton structures ranging from single-hump to double-hump and flat-top profiles \cite{tkjpa} that results due to the contribution from four-wave mixing nonlinearity of the system (\ref{gen-2ccnls}). However, the present soliton solution can be considered as a special case of that one \cite{tkjpa} with the same dynamics on both components, namely a degeneracy state ($q_1$ and $q_2$ are identical or simply $q_1=q_2$). For completeness, we have depicted such stable degenerate solitons in Fig. \ref{fig-soliton}(a). Thus, in some sense, the present switching parameter $s$ can be associated/equivalent with/to the auxiliary function given in Ref. \cite{tkjpa}. {Here, the entire spectrum of wave structures and their dynamics can be controlled by a single parameter `$s$' (relative to `$a$' ). Based on the dynamics, it can be associated with phase-dependent coherent nonlinearity of the system which shows promising energy-switching behaviour  \cite{tkjpa}.} For a more detailed understanding of the general bright solitons and their collisions, one can refer \cite{tkjpa}. Importantly, the ramifications of varying nonlinearities in such solitons and their collisions shall also be the next immediate assignment along this direction. 

As the main objective of this work is on the inhomogeneity, we come back to the general solution (\ref{sol-solution}) and consider the modulated nonlinearities given in Eq. (\ref{nonlinearity}). The periodic nonlinearity modulation $\sigma_1(z)$ transforms the stable soliton into a breather, which oscillates in the amplitude and along the $z$ direction. However, there is no significant change along $t$, which can be visualized from Fig. \ref{fig-soliton}(b). This can be viewed as Ma type breather oscillating along propagation direction but with zero background in contrary to the standard Ma breathers, which usually appear on a non-zero constant background. In addition to $a$, $\epsilon_1$ and $\epsilon_2$ values, one can control the amplitude, width and velocity of these breathing solitons by tuning the arbitrary parameters $b_0$, $b_1$, $b_2$, and $b_3$.  

Moving further to the case of step-like nonlinearity $\sigma_2=b_0+b_1 ~\text{tanh}(b_2 z+b_3)$ (\ref{nonlinearity}b), we observe an interesting phenomenon called  amplification accompanied with a compression which can also be visualized as selective amplification through cascaded compression. This is a single-step steady increase (decrease) in the amplitude (width) of the soliton, see Fig. \ref{fig-soliton}(c). A reverse transition can also be achieved by suppressing the energy/intensity of a pulse/beam along with a suitable increase in bandwidth when $b_1<0$. Also, by controlling the nonlinear phase factor $b_3$ we can shift the critical transition point along `$z$' based on the requirement. 

The bell type (`sech') nonlinearity $\sigma_3(z)$ induces a localized change in the available wave of any characteristics. This can be broadly of two types; the first one has localized excitation or exciton formation in the absence of background of the bell nonlinearity ($b_0=0$). The second consequence of the bell nonlinearity is a tunneling process characterized by its amplitude parameter $b_1$. It can be further divided into two types, namely the tunneling-through a high potential barrier (barrier penetration for $b_1>0$) and cross-over of a potential well ($b_1<0$). In the earlier exciton formation process, the stable soliton switches into a rogue wave-type structure localized in $z$ with extended tails along $t$ as shown in Fig. \ref{fig-soliton}(d). Such a localized modulation resembles the collision scenario of two oppositely moving bright solitons, which produces a maximum intensity in the collision regime. Additionally, in the case of tunneling and cross-over processes, the solitons appear stable before and after the barrier/well without any other change. At the potential barrier/well, they induce a localized hump/dip and we refrain from giving its graphical demonstration here. Beyond the limited functionalities of the present solitons, the impression of nonlinearities in other localized structures is much more interesting, as shown in the forthcoming discussions. 

\subsection{Inhomogeneous Akhmediev Breathers}
Beyond the standard solitons discussed above, solution (\ref{solution}) exhibit periodically oscillating structures called breathers for $s\neq 0$. One such entity is Akhmediev breathers, which is nothing but a nonlinear wave localized along the propagation direction $z$ and periodic along the transverse direction $t$. This arises for the choice $a,s\neq 0$ and $\tau^2 <0$ in Eq. (\ref{solution}) and the corresponding mathematical expression can be rewritten as below.
\bes \bea 
&&q_1= \epsilon_1 \sqrt{\sigma}  ~\frac{(a^2-s^2)~ \mbox{cosh}[4 a b ~Z] + i a b ~\mbox{sinh}[4 a b~ Z]}{a~ \mbox{cos}[2 b~ T] - s~ \mbox{cosh}[4 a b~ Z]}e^{i(\xi-2 s^2 Z)},\\
&&q_2=\epsilon_1 \sqrt{\sigma} ~\frac{a^2~ \mbox{cosh}[4 a b ~Z] + i a b ~\mbox{sinh}[4 a b~ Z]-s a~ \mbox{cos}[2 b~ T]}{a~ \mbox{cos}[2 b~ T] - s~ \mbox{cosh}[4 a b~ Z]}~e^{i(\xi-2 s^2 Z)},
\eea \label{AB-solution} \ees
where $b^2=|a^2-s^2|$, $\xi$, $T$ and $Z$ take the form of as given in Eq. (\ref{cons}). Interestingly, as mentioned in cases (ii) and (iii) of the Table \ref{table1}, one can obtain bright, gray and dark type Akhmediev breathers with respect to different choices of $a$ and $s$ parameters. Further note that the $q_1$ and $q_2$ components admit single-hump and double-well (or double-dip) breathing patterns, respectively. Also, their intensity/amplitude shall be tuned by varying these $a$ and $s$ in addition to the transformation parameters. 

\begin{figure}[h]
	\centering\includegraphics[width=0.25\linewidth]{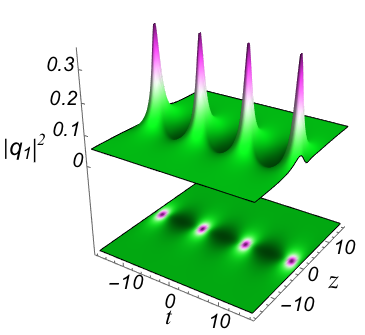}\includegraphics[width=0.25\linewidth]{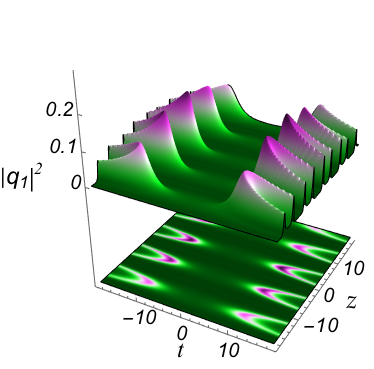}\includegraphics[width=0.25\linewidth]{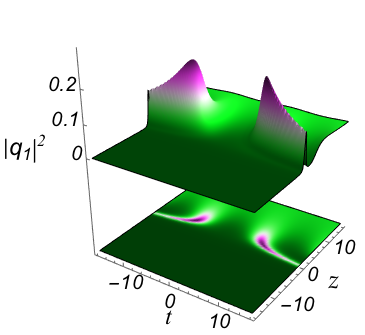}\includegraphics[width=0.25\linewidth]{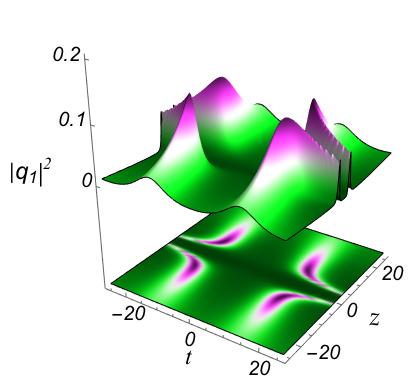}\\%\includegraphics[width=0.25\linewidth]{AB-sech-q1}\\
	\centering\includegraphics[width=0.25\linewidth]{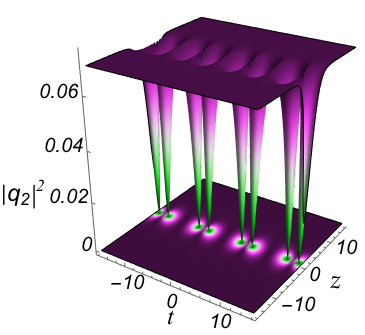}\includegraphics[width=0.25\linewidth]{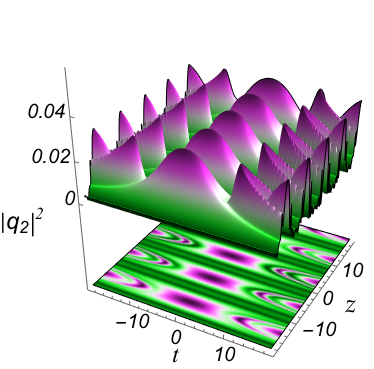}\includegraphics[width=0.25\linewidth]{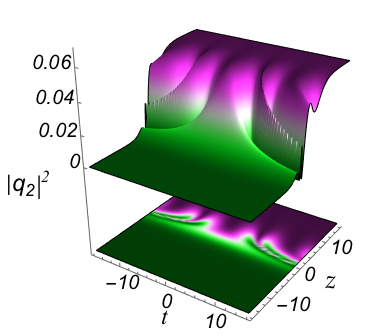}\includegraphics[width=0.25\linewidth]{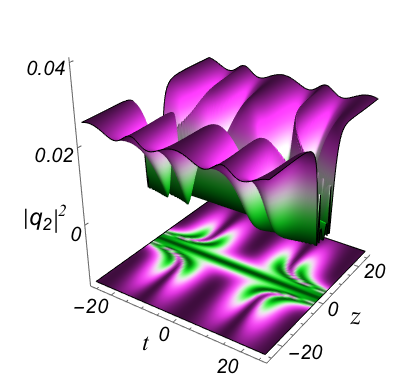}\\%\includegraphics[width=0.25\linewidth]{AB-sech-q2}
	{\hfill\qquad (a) \hfill\qquad (b) \hfill \qquad (c) \hfill \qquad\qquad(d) \hfill} 
	\caption{Dynamics of bright-dark Akhmediev breathers for $s=1$ and $a=-0.75$ with (a) constant nonlinearity and their transformation due to (b) periodic, (c) kink-like, and (d) bell-type nonlinearities revealing (b) localization-broken doubly-periodic breathers, (c) escalated background amplitude time-periodic breathers, and (d) localization retaining centrally symmetric breathers, respectively. The parameters are chosen as (a) $b_0 = 2.0$, $b_1 =b_2 =b_3= 0.0$, (b) $b_0 = 0.4$, $b_1 = 0.5$, $b_2=0.75$ \& $b_3 =0.02$; (c) $b_0 = 0.75$, $b_1 = 0.75$, $b_2=0.5$ \& $b_3 =0.02$, and (d) $b_0 = 0.75$, $b_1 = -0.75$, $b_2 = 0.5$, \& $b_3 =0.02$ with other values fixed as $\epsilon_1 = 0.25$, and $\epsilon_2 = 0.05$. } \label{fig-AB}
\end{figure}

%\rhorally-localized--spatially-periodic 
The obtained Akhmediev breathers can be controlled with the help of varying nonlinearities and it modulates their identities like localization, periodicity, amplitude, width/thickness of the beam, etc. Particularly, with a periodically varying nonlinearity $\sigma_1(z)$, its localization is broken and now becomes as doubly-periodic breathers, which means periodic in both $z$ and $t$. %\rhorally and spatially periodic. 
To be precise, in the $q_1$ component, the single-hump excitations split/deform into soliton lattices on either side of the central dip and this form periodically manifest along $z$ indefinitely with substantially lesser amplitude. On the other hand, in $q_2$ component, the double-well with central maximum intensity spread out by breaking the localization and also become periodic along `$z$' as shown in Fig. \ref{fig-AB}(b). Here the important ramification of periodic nonlinearity in Akhmediev breathers is on the localization and amplitude. Also, the period of oscillation and amplitude can be altered by tuning the arbitrary $b_j$ as well as $\epsilon_j$ parameters.  

Contrary to the periodic nonlinearity, the kink-like (tan-hyperbolic) nonlinearity $\sigma_2$ (\ref{nonlinearity}b) does not alter/break the localization. Still, the modulation preserves the localization in $z$ with a centrally-symmetric deformation and their background energies are uniformly escalated like a step function in both components. Also, unlike in the case of solitons, here the kink nonlinearity does not induce any amplification or compression of the localized excitations, and we have shown such a kink modulated breather in Fig. \ref{fig-AB}(c). 

Another kind of hyperbolic nonlinearity $\sigma_3=b_0+b_1~ \mbox{sech}(b_2 z+b_3)$ having bell-type localized form also not affect the localization of the Akhmediev breather in the present system. However, it alters the nature of periodic structures from uniform breathing to a centrosymmetric double-peaked breather, as shown in Fig. \ref{fig-AB}(d) for one class of nonlinearity parameters $b_j\neq 0$. For a different set of parameters ($b_0 =0$) one can witness the algebraically localized rogue-wave-type excitons with sustaining side-band tails along $z$. Additionally, the single-hump and double-dip nature of the Akhmediev breathers are substantially preserved. 

\subsection{Inhomogeneous Ma Breathers}
\begin{figure}[h]
	\centering\includegraphics[width=0.25\linewidth]{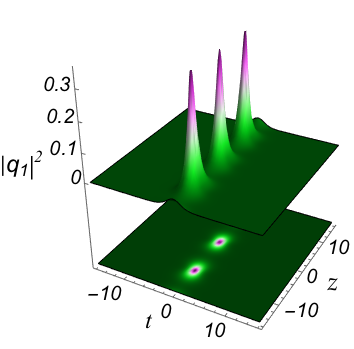}\includegraphics[width=0.25\linewidth]{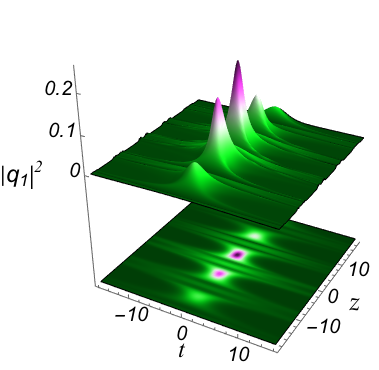}\includegraphics[width=0.25\linewidth]{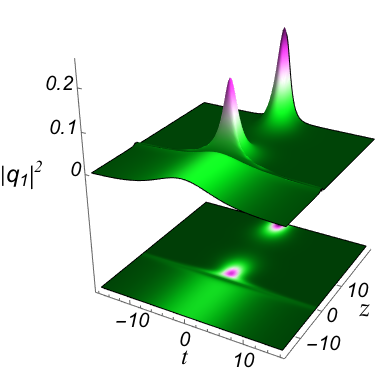}\includegraphics[width=0.25\linewidth]{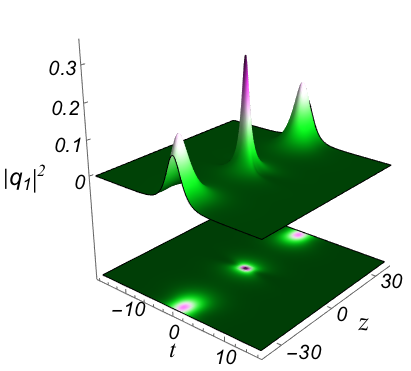}\\%\includegraphics[width=0.25\linewidth]{MB-sech-q1-b}\\
	\centering\includegraphics[width=0.25\linewidth]{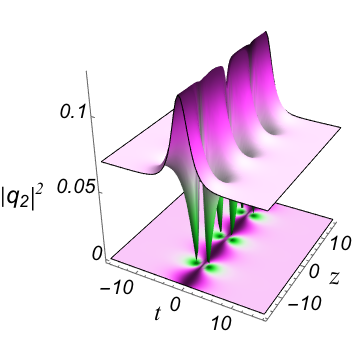}\includegraphics[width=0.25\linewidth]{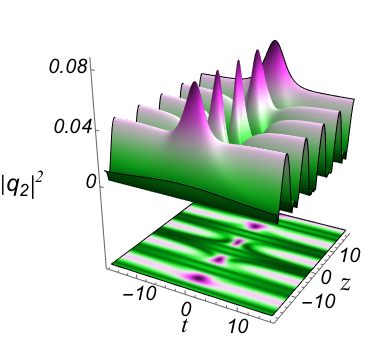}\includegraphics[width=0.25\linewidth]{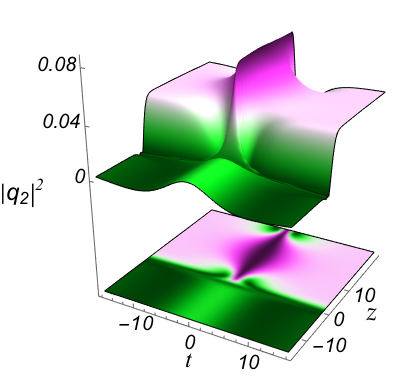}\includegraphics[width=0.25\linewidth]{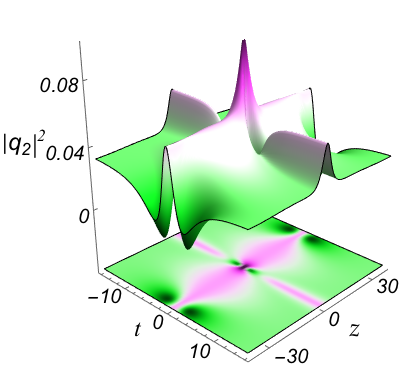}\\%\includegraphics[width=0.25\linewidth]{MB-sech-q2-b}
	{\hfill\qquad (a) \hfill\qquad (b) \hfill \qquad (c) \hfill \qquad\qquad(d) \hfill} 
	\caption{Dynamics of modulated Ma breathers for $s=0.75$ and $a=1.0$ with (a) constant, (b) periodic, (c) kink-like, and (d) bell-type nonlinearities projecting (b) localization-broken, (c) localization-sustaining escalated amplitude, and (d) localization-preserving centrally excited breathers, respectively. The parameters are chosen as (a) $b_0 = 2.0$, $b_1 =b_2 =b_3= 0.0$, (b) $b_0 = 0.5$, $b_1 = 0.75$, $b_2=0.75$ \& $b_3 =0.02$; (c) $b_0 = 0.5$, $b_1 = 0.75$, $b_2=0.75$ \& $b_3 =0.02$, and (d) $b_0 = 0.95$, $b_1 = 0.75$, $b_2 = 0.5$, \& $b_3 =0.02$ with other values fixed as $\epsilon_1 = 0.25$, and $\epsilon_2 = 0.05$. } \label{fig-MB}
\end{figure}

Opposite to the Akhmediev breathers given in the previous part, On the other hand, for $\tau^2 >0$, one can have another form of breathing structure from the general wave solution (\ref{solution}). They are periodic along the propagation direction $z$ and localized in the transverse coordinate $t$ and are usually referred as Ma breather or Kuznetsov-Ma soliton. In a mathematical sense for this choice, Eq. (\ref{solution}) reduces to the following simple form:
\bes \bea 
&&q_1=\epsilon_1 \sqrt{\sigma}~ \frac{(s^2-a^2)~ \mbox{cos}[4 a b ~Z] + i  a b~ \mbox{sin}[4 a b~ Z]}{s~ \mbox{cos}[4 a b ~Z] - a~ \mbox{cosh}[2 b~ T]}~ e^{i(\xi-2 s^2 Z)},\\
&&q_2=\epsilon_1 \sqrt{\sigma} ~\frac{s a~ \mbox{cosh}[2 b~ T] - a^2~ \mbox{cos}[4 a b ~Z] + i a b~ \mbox{sin}[4 a b~ Z]}{s~ \mbox{cos}[4 a b ~Z] - a~ \mbox{cosh}[2 b~ T]}~e^{i(\xi-2 s^2 Z)},
\eea \label{MB-solution} \ees
where the form of $\xi$, $T$, and $Z$ are as defined in Eq. (\ref{cons}) while $b^2=|a^2-s^2|$. Similar to the Akhmediev breathers, the Ma breathers also exhibit both bright-dark and dark-dark type single-hump/single-well and double-well localized structures for appropriate choices of $s$ and $a$ as shown in Table \ref{table1}. Still, the localization and periodic directions get exchanged here. For illustrative purposes, we portray bright-dark Ma breathers in Fig. \ref{fig-MB}(a) for the constant nonlinearity $\sigma(z)=2$..

The above given Ma breathers can also be tailored by every type of varying nonlinearities given in \ref{nonlinearity}. To start with, the periodic nonlinearity $\sigma_1=b_0+b_1~ \mbox{sin}(b_2 z+b_3)$ modulates the background of the breathers by introducing oscillations and redistributes their peak intensity. The newly developed background oscillations continuously decrease and will vanish away as $t\rightarrow \pm \infty$ in the $q_1$ component. However, in the $q_2$ component, those background oscillations are stable and sustain even when $t\rightarrow \pm \infty$  due to the non-zero initial intensity in dark or gray breathers as evidenced from Fig. \ref{fig-MB}(b). Another feature is that the peak intensity oscillations repeat periodically along the propagation direction $z$. As usual, the arbitrary parameters $b_0$, $b_1$, $b_2$, $b_3$, $\epsilon_1$ and $\epsilon_2$ do help in manipulating the obtained Ma breathers.

Next, the kink-like nonlinearity acts as a simple intensity amplifier with cascaded compression in Ma breathers and is likely to be similar to the behaviour observed in inhomogeneous solitons (\ref{sol-solution}). Here the breathing structures reemerge after a step-amplification from an initial stable profile without any change in their localization $t$ and retain their periodic oscillation of intensity along $z$ in both components for bright as well as dark Ma breathers, refer Fig. \ref{fig-MB}(c). Here the period of breathing oscillations and amplitude of peaks/dips get modulated and they can be tuned by using $b_j$, and $\epsilon_j$ parameters. The third type of nonlinearity, bell-type $\sigma_3$, displays all the three effects in the Ma breathers as well starting from the tunneling ($b_1>0$), cross-over ($b_1<0$) and localized exciton formation ($b_0=0$) based on the $b_j$, $j=0,1,2,3$ parameters. For the demonstration, we have shown the tunneling effect in Fig. \ref{fig-MB}(d), which preserves the uniform breathing patterns with a modified/increased period of oscillations. It induces a localized maximum intensity at the barrier in both components, while it generates long-lasting sideband tails only in the dark $q_2$ mode. Similarly, the cross-over and exciton formation can also be observed, which also preserves the nature of breathers with a substantial change in the amplitudes. 

\subsection{Inhomogeneous Rogue Waves}
\begin{figure}[h]
	\centering\includegraphics[width=0.235\linewidth]{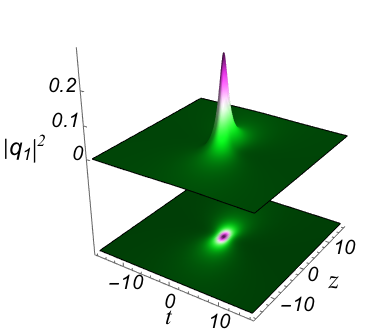} \includegraphics[width=0.235\linewidth]{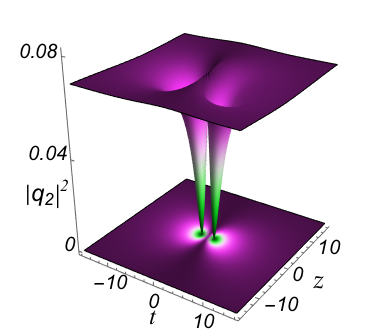}~
	\includegraphics[width=0.235\linewidth]{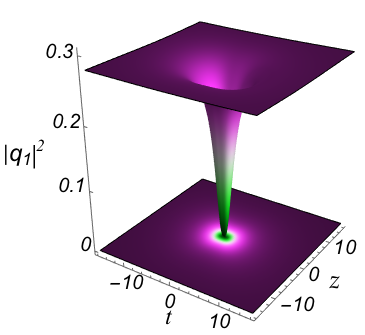}\includegraphics[width=0.235\linewidth]{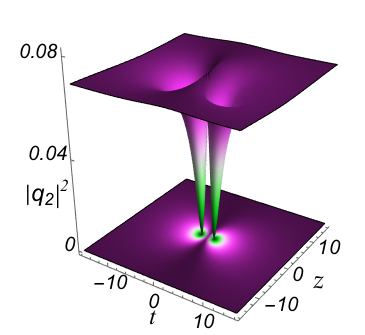}\\
	{\hfill\qquad (a) \hfill\qquad (b) \hfill \qquad (c) \hfill \qquad\qquad(d) \hfill}
	\caption{Dynamics of (a-b) bright-dark rogue waves for $s=0.75$ and (c-d) dark-dark rogue waves for $s=-0.75$ for constant nonlinearity $\sigma=2$ with $a=0.751$. } \label{fig-rogue-bd}% to which $\tau$ becomes equal but the factor $s/a$ is positive and negative, respectively 
\end{figure}

{Rogue waves are doubly-localized (both in space and time or both in propagation and transverse directions) structures with extreme amplitudes over a continuous/constant background. It is evident from solution (\ref{solution}) that the present CCNLS system (\ref{gen-2ccnls}) with modulated nonlinearity admits both bright and dark rogue waves of single-hump/single-dip and double-dip profiles for a particular choice of parameters `$|s|\approx |a|$' as shown in the Table \ref{table1} and as demonstrated in Fig. \ref{fig-rogue-bd}}. One can deduce an explicit mathematical form for rogue waves from either Akhmediev breathers (\ref{AB-solution}) or Ma breathers (\ref{MB-solution}) with a condition that $\tau \rightarrow 0$ (and subsequently $b\rightarrow 0$). One main difference in the present solution is that the rogue waves do not have side-band tails, which makes it look like standard lumps/wells. Also, the amplitude/depth of the bright/dark rogue wave shall be controlled by the parameter $s$.

\begin{figure}[h]
	\centering\includegraphics[width=0.25\linewidth]{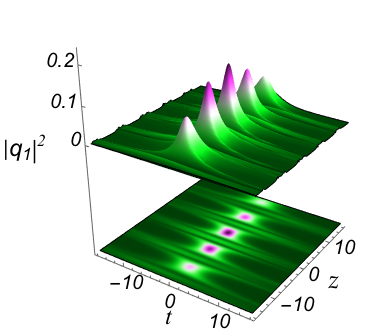}\includegraphics[width=0.25\linewidth]{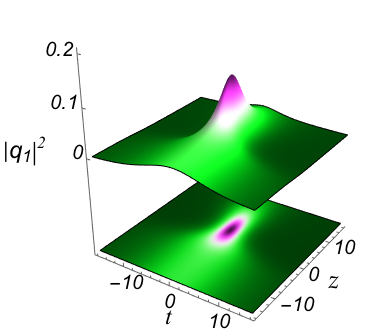}\includegraphics[width=0.25\linewidth]{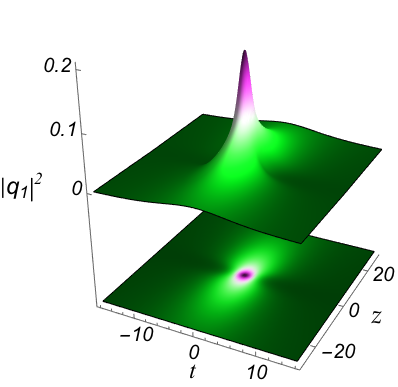}\includegraphics[width=0.25\linewidth]{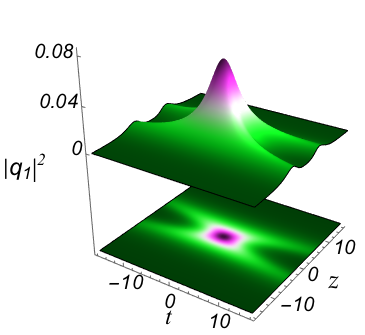}\\
	\includegraphics[width=0.25\linewidth]{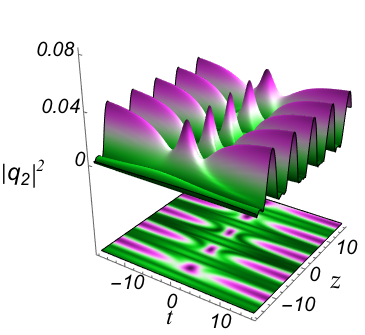}\includegraphics[width=0.25\linewidth]{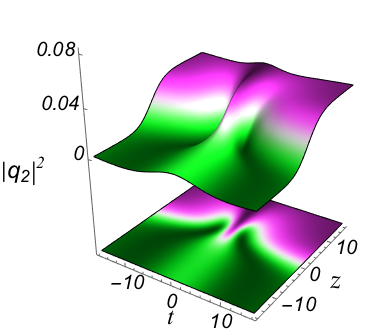}\includegraphics[width=0.25\linewidth]{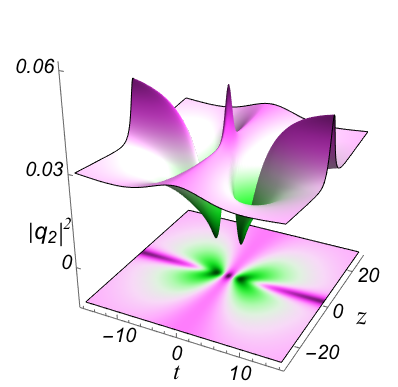}\includegraphics[width=0.25\linewidth]{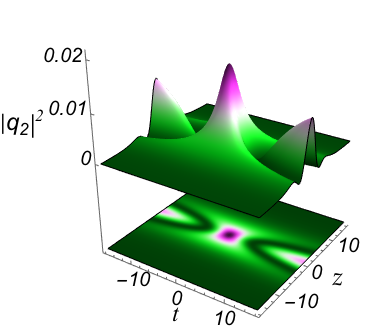}\\ {\hfill\qquad (a) \hfill\qquad (b) \hfill \qquad (c) \hfill \qquad\qquad(d) \hfill} 
	\caption{Dynamics (a) A-shaped \& M-shaped periodic wave trains, (b) localized amplification with compression, (c) tunneling, (d) exciton with side band formation of bright-dark rogue waves due to modulated (a) periodic, (b) kink-like, and (c-d) bell-type nonlinearities for $s=0.75$ and $a=0.7501$. Other parameters are chosen as (a) $b_0 = 0.5$, $b_1 =0.75$, \& $b_2 =0.5$; (b) $b_0 = 0.75$, $b_1 = 0.5$, \& $b_2=0.25$; (c) $b_0 = 0.75$, $b_1 = 0.25$, \& $b_2=0.5$, and (d) $b_0 = 0.0$, $b_1 = 0.5$, \& $b_2 = 0.5$ with $b_3 =0.02$, $\epsilon_1=0.25$ and $\epsilon_2=0.05$. } \label{fig-rogue}
\end{figure}

As the name suggests, the periodically varying nonlinearity of the form $\sigma_1=b_0+b_1 ~\text{sin}(b_2 z+b_3)$ breaks the doubly localized rogue waves into periodic one akin to breathers. Especially, the zero-background bright rogue wave admits a train of peaks with high amplitude in the center and decreases in either direction along $z$. Along the $t$ direction, initially, there appear small-amplitude tails which further decrease and finally vanish away. Instead of a single-hump, the nonlinearity modulates it to exhibit multi-periodic A-shaped humps and vanishing thereafter. On the other hand, in dark rogue wave ($q_2$ component), it induces periodic $W$-shaped structures around the center. As $z$ increases, the $W$-shape becomes a constant amplitude and then periodic stable waves as shown in Fig. \ref{fig-rogue}(a). In one way, this may be similar to a periodic modulation of the Ma breathers, but in the earlier case, it is periodically repeating, and here it is not so. Further, by tuning the parameters $a$, $s$, $b_j$, and $\epsilon_j$, the periodicity, amplitude, and width shall be manipulated appropriately.  

The kink-like nonlinearity $\sigma_2=b_0+b_1~ \mbox{tanh}(b_2 z+b_3)$ given by Eq. (\ref{nonlinearity}b) leads to initial widening of the doubly-localized rogue waves with lesser amplitude and focuses it around the switching/amplification regime leading to enhanced intensity in both components. Then both localized wave intensity as well as the background energy got increased and remain stable throughout without any emergence of periodic structures, as portrayed in Fig. \ref{fig-rogue}(b). It is very clear that the localization is preserved after the amplification of both bright and dark rogue waves. Compared to this kink nonlinearity, the bell-type nonlinearity shows more promising evolution which includes the already introduced tunneling Fig. \ref{fig-rogue}(c) and exciton formation Fig. \ref{fig-rogue}(d) in addition to the cross-over effect. As these effects are well described in the previous cases, here we refrain from repeating the same discussion. In a general picture, it seems that the modulation due to nonlinearities in the Ma breathers and the rogue waves similar. However, more careful analysis brings out the differences between them. Especially, the nature of modulated waveforms is different in both cases except for the respective localization preservation. Further, these two localized waves show a completely different pattern of modulations compared to that of the Akhmediev breathers. 

\subsection{Importance of the Similarity Parameters}
\begin{figure}[h]
	\centering\includegraphics[width=.25\linewidth]{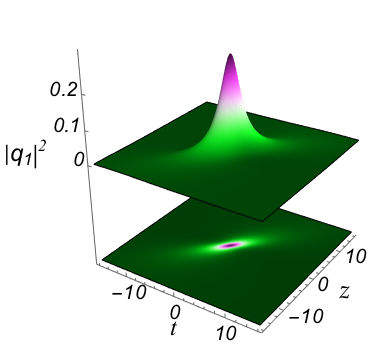}~\includegraphics[width=.25\linewidth]{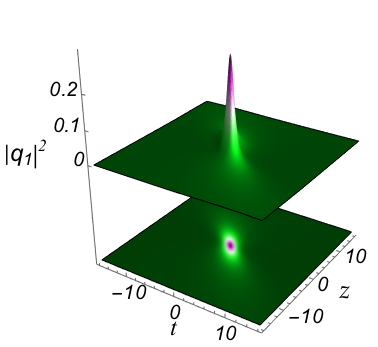}\includegraphics[width=.25\linewidth]{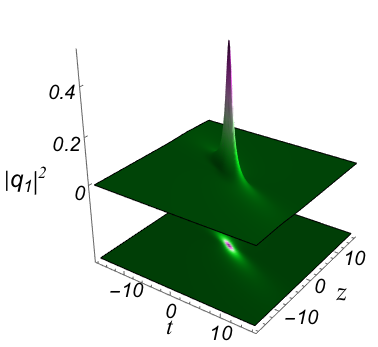}~\includegraphics[width=.25\linewidth]{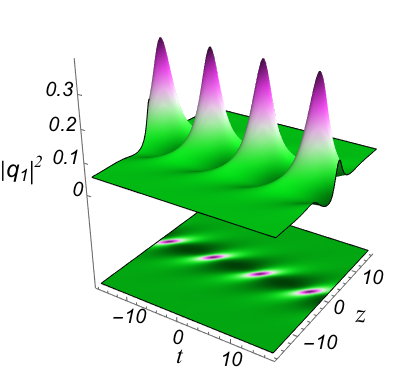}\\
	\includegraphics[width=.25\linewidth]{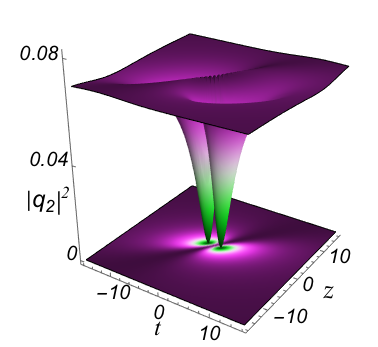}~\includegraphics[width=.25\linewidth]{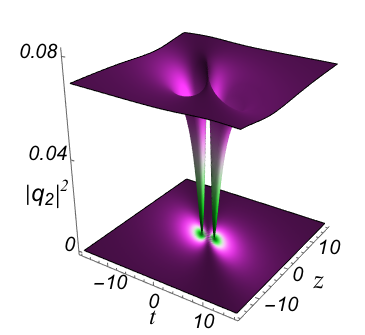}\includegraphics[width=.25\linewidth]{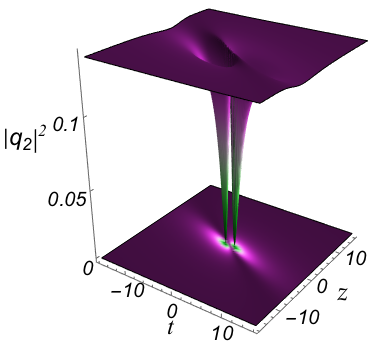}~\includegraphics[width=.25\linewidth]{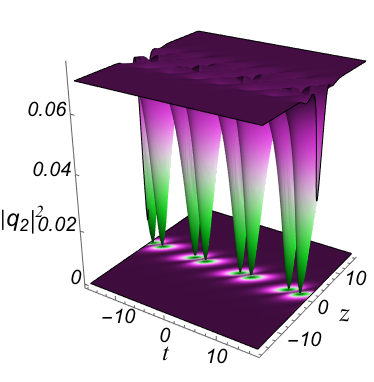}\\ {\hfill\qquad (a) \hfill\qquad (b) \hfill \qquad (c) \hfill \qquad\qquad(d) \hfill} 
	\caption{Change in the inclination/angle of single-hump bright and double-well dark rogue waves for different $\epsilon_1$ and  $\epsilon_2$ parameters. Their values are (a) $\epsilon_1=0.25$ and  $\epsilon_2=-2.75$, (b) $\epsilon_1=0.25$ and  $\epsilon_2=2.75$, and (c) $\epsilon_1=0.35$ and $\epsilon_2=2.75$ while the remaining parameters are as given in Fig. \ref{fig-rogue}. (d) Change in the angle of gray-dark Akhmediev breathers for $\epsilon_1=0.25$ and  $\epsilon_2=-3.75$  with other parameters as $a=1.0$, $s=-0.75$, $b_0=2$ and $b_j=0,~j=1,2,3$, but their localization remain undisturbed.}
	\label{fig-rogue-angle}
\end{figure}

The above discussions have provided a deeper insight into the role of various modulated nonlinearities and their significance on the nonlinear coherent structures in the system (\ref{gen-2ccnls}). To be specific, these modulated nonlinearities that can be tuned suitably by $b_j$ parameters influence the stationary solitons as well as bright, dark/gray, bright/gray-dark/gray type Akhmediev breathers, Ma breathers, and rogue waves. Still, some interesting hidden factors influencing these coherent structures are yet to be addressed. An important point to be noted here is that the significance of the two arbitrary parameters $\epsilon_1$ and  $\epsilon_2$ available under the similarity transformation. Particularly, $\epsilon_2$ provides the possibility of determining the inclination/angle of the localized structures without altering any of their other identities. But, $\epsilon_1$ parameter supports the change of inclination at a very small level compared to that of $\epsilon_2$. However, the change in $\epsilon_1$ highly affects the amplitude and helps in making the wave to a desirable width, which is inversely proportional to the amplitude. So, one can readily transform the rogue waves to a wider (or narrower) one with smaller (or larger) amplitude based on the requirement. For illustrative purposes, we have shown the change in the inclination of bright-dark rogue waves in Fig. \ref{fig-rogue-angle}(a-c). A similar property can be observed in the Akhmediev breathers, which remain localized in $z$ and induces inclination/angle of the single-hump/double-well structures, as evidenced from Fig. \ref{fig-rogue-angle}(d). 

\begin{figure}[h]
	\centering\includegraphics[width=.25\linewidth]{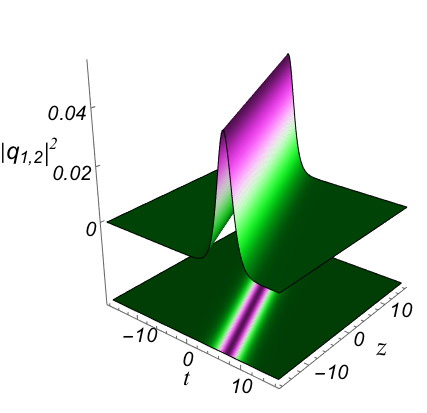}~\includegraphics[width=.25\linewidth]{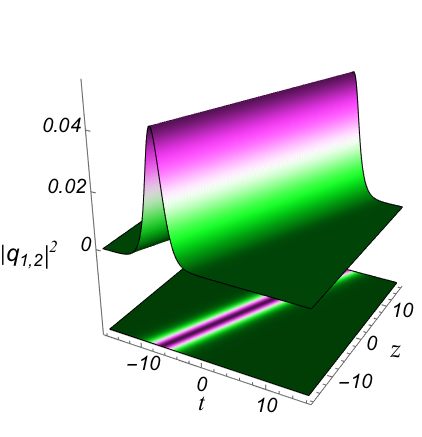}\includegraphics[width=.25\linewidth]{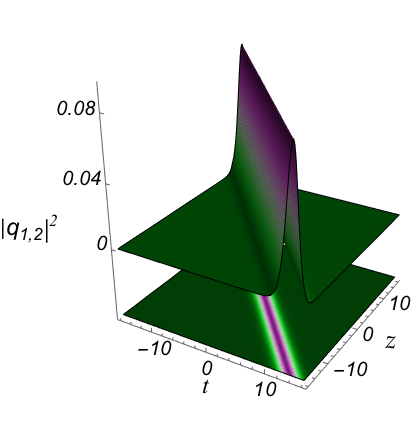}\\{\hfill\hfill (a) \hfill \hfill (b) \hfill \hfill (c) \hfill\hfill} 
	\caption{Traveling degenerate bright solitons for different choices of $\epsilon_j$ parameters with $s=0$, $a=0.6$, $b_0=2$ and $b_1=b_2=b_3=0$. (a) Left moving solitons for $\epsilon_1=0.25$ and $\epsilon_2=1.75$, (b) Right moving solitons for $\epsilon_1=0.25$ and $\epsilon_2=-1.75$, and (c) Intensity increasing and compressed soliton with velocity change for $\epsilon_1=0.35$ with $\epsilon_2=1.75$. }
	\label{fig-soliton-angle}
\end{figure}

\begin{figure}[h]\vspace{-0.3cm}
	\centering\includegraphics[width=.25\linewidth]{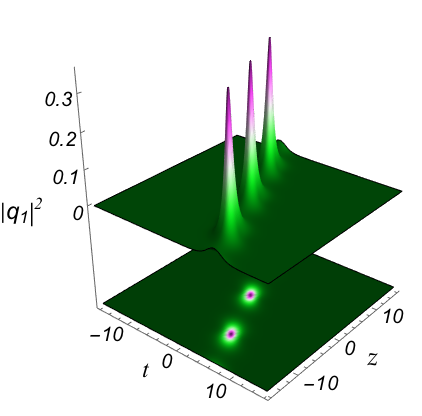}~\includegraphics[width=.25\linewidth]{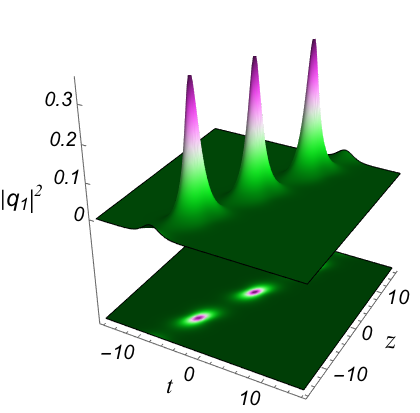}\includegraphics[width=.25\linewidth]{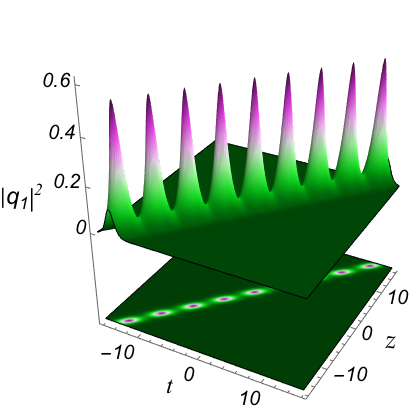}\\
	\centering\includegraphics[width=.25\linewidth]{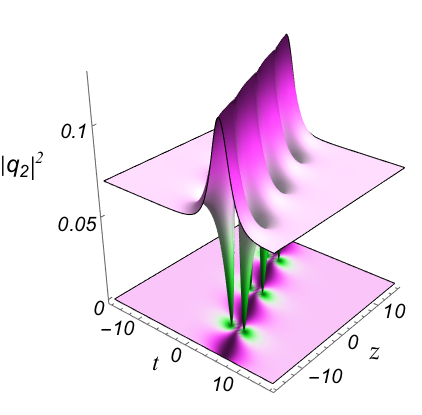}~\includegraphics[width=.25\linewidth]{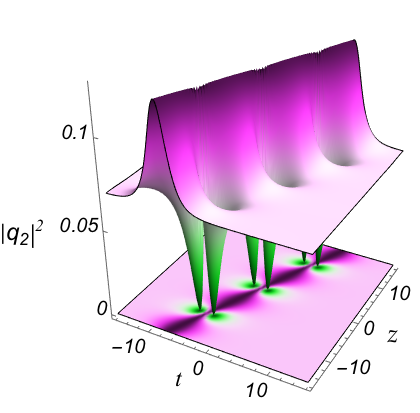}\includegraphics[width=.25\linewidth]{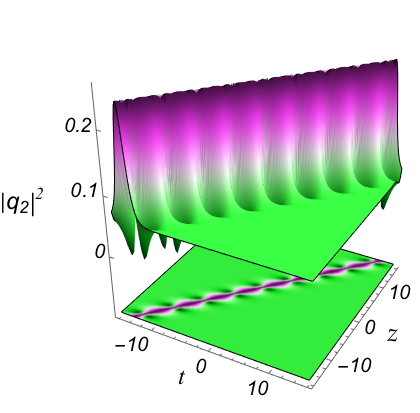}\\{\hfill\hfill (a) \hfill \hfill (b) \hfill \hfill (c) \hfill} 
	\caption{Transforming bright-dark type Ma breathers into traveling general (non-localized) breathers for different choices of $\epsilon_j$ parameters. (a) Left moving breathers for $\epsilon_2=0.25$ and $\epsilon_2=1.75$, and (b) Right moving breathers for $\epsilon_2=0.25$ and $\epsilon_2=-1.75$, (c) Right moving general breathers with Intensity/oscillation increase and compression with a small change in the velocity for $\epsilon_1=0.35$ with $\epsilon_2=-1.75$. Other parameters are chosen as $a=0.75$, $s=1.0$, $b_0=2$ and $b_1=b_2=b_3=0$. }
	\label{fig-MB-angle}
\end{figure}

Interestingly, by utilizing these $\epsilon_j$ parameters, we can transform the stationary solitons to the traveling (left or right moving) solitons with appropriate amplification of intensity accompanied by compression/expansion and they influence the central position and velocity of the associated solitons too. We have depicted such propagating degenerate solitons in Fig. \ref{fig-soliton-angle} for further insight. Further, from the $t$-localized Ma breathers, one can also obtain a set of generalized breathers that are neither localized in propagation direction not transverse direction, and they propagate with specific velocity defined by the parameters. To be precise, localization-breaking ($t$-localized to non-localized) breathers are achieved by $\epsilon_1$ and $\epsilon_2$ parameters, which we have demonstrated such breather phenomenon in Fig. \ref{fig-MB-angle} for a better understanding. 
Apart from the above, the explored features such as velocity-controlled solitons and breathers, changing the angle of the peak/hole intensities rogue waves and Akhmediev breathers, intensity as well as periodicity tailoring in solitons and Ma breathers, shall be effectively employed in the respective type of above discussed modulated nonlinearities by tuning $\epsilon_1$ and $\epsilon_2$. 
 

As a future study, the present investigation shall be extended to general solitons and their collisions with appropriate modulated nonlinearities which is reported recently \cite{jpa20}. Further, one can also study the dynamics of various nonlinear coherent structures, their co-existence and dynamics in dispersion as well as nonlinearity management in various systems, for example \cite{tkscr19,He18}, by incorporating gain/loss too. Possible outcomes will be reported separately with categorical analysis.
%\newpage
\section{Conclusions}\label{sec-conclusion}
In this work, we have considered a coherently coupled nonlinear Schr\"odinger system consisting of modulated self-phase modulation, cross-phase modulation, and four-wave mixing nonlinearities with a varying refractive index. We have obtained a general two-parameter localized wave solution by using an appropriate similarity transformation and following the Ref. \cite{pre2015}. By adopting three different types of varying nonlinearities, namely periodic, kink-like, and bell-type functions, we have explored the manipulation mechanism of the resultant nonlinear wave structures. Notably, we found that through proper engineering of parameters, the present solution manifests into various forms of localized waves ranging from stable degenerate solitons to Akhmediev breathers, Ma breathers, and doubly-localized rogue waves of the bright, gray and dark type. Additionally, the modulated nonlinearities revealed various phenomena such as periodically oscillating waves (from localized to periodic waves), amplification of localized waves with cascaded compression, tunneling as well as cross-over of the excited barrier/well, and localized exciton formation with sideband tails. Further, the controlling mechanism of individual identities of each localized waves like velocity, amplitude, central position, and orientation/angle/inclination, in addition to localization-breaking features resulting in general breathers from Ma breathers are possible with the considered type of varying nonlinearity parameters. The results presented in this work will be applicable to the studies on engineering localized waves like solitons, breathers, and rogue waves as well as to various experimental investigations on the controlling mechanism of rogue waves in optical systems, cavity soliton based fiber lasers, intracavity and pump waves in polarization controlled waveguides, atomic condensates, deep water oceanic waves, and other related coherent wave systems that can be implemented with nonlinearity management.\\ 

\setstretch{1.0}
\noindent{\bf Acknowledgement}: The work of KS was supported by Department of Science and Technology - Science and Engineering Research Board (DST-SERB), Govt. of India, sponsored National Post-Doctoral Fellowship (File No. PDF/2016/000547). %KS is grateful to Prof.M. Lakshmanan, Professor of Eminence and DST-SERB Distinguished Fellow, Centre for Nonlinear Dynamics, Bharathidasan University, Tiruchirappalli, India, for his constant support and encouragement. 
The work of TK was supported by the DST-SERB, Govt. of India, in the form of a Major Research Project (File No. EMR/2015/001408).
%\newpage

%\section*{References}

\end{document}